\documentclass[a4paper,pre,aps,showpacs,byrevtex,amsmath,amssymb,onecolumn,nofootinbib]{revtex4}
\usepackage{graphicx}
\usepackage{amsmath}
\usepackage{amsfonts}
\usepackage{amssymb}

\newcommand{\tv}{\mathbf{t}}

\newcommand{\fo}{\mathcal{F}}
\newcommand{\de}{\delta}
\newcommand{\En}{\mathcal{E}}
\newcommand{\df}[1]{\overset{.}{#1}}
\newcommand{\ddf}[1]{\overset{..}{#1}}
\newcommand{\dddf}[1]{\overset{...}{#1}}
\newcommand{\vev}[1]{ \left\langle #1 \right\rangle}
\DeclareMathOperator{\sn}{sn}
\DeclareMathOperator{\cn}{cn}
\DeclareMathOperator{\dn}{dn}
\DeclareMathOperator{\am}{am}
\DeclareMathOperator{\E}{E}

\DeclareMathOperator{\K}{K}
\DeclareMathOperator{\Hyp}{\Phi}
\DeclareMathOperator{\F}{F}
\DeclareMathOperator{\Lam}{\Lambda}

\DeclareMathOperator{\gam}{\Gamma}
\begin{document}

\title{Semi-Classical Buckling of stiff polymers}

\author{Marc Emanuel$^1$, Herv\'{e} Mohrbach$^2$, Mehmet Sayar$^3$, Helmut Schiessel$^1$, and Igor M. Kuli\'c$^4$}

\address{$^1$Instituut-Lorentz, Universiteit
Leiden, Postbus 9506, 2300 RA Leiden, The Netherlands\\
$^2$Institut de Physique, Universit\'e Paul Verlaine-Metz, LPMC, CPMB1-FR
CNRS 2843, 1 boulevard Arago, 57078 Metz, France\\
$^3$Koc University, College of Engineering, 34450 Sariyer, Istanbul, Turkey \\
$^4$School of Engineering and Applied Sciences, Harvard University, Cambridge, Massachusetts 02138, USA}

\begin{abstract}
A quantitative theory of the buckling of a worm like chain based on a semi-classical approximation of the partition function is presented. The contribution of thermal fluctuations to the force-extension relation that allows to go beyond the classical Euler buckling is derived in the linear and non-linear regime as well. It is shown that the thermal fluctuations in the nonlinear buckling regime increase the end-to-end distance of the semiflexible rod if it is confined to 2 dimensions as opposed to the 3 dimensional case.
Our approach allows a complete physical understanding of buckling in $D=2$ and in $D=3$ below and above the Euler transition.

\end{abstract}

\maketitle

\section{Introduction}
During the last few years, the advent of single molecule nanomanipulation \cite{Strick2003} has allowed to study the elastic properties of DNA and other biopolymers under different physical conditions. In these experiments, the extension of single molecule 
versus an applied stretching force is measured by a variety of techniques including magnetic beads \cite{Smith1992,Strick1996}, optical traps \cite{Smith1996,Wang1997}, micro-needles \cite{Cluzel1996}, hydrodynamic flow \cite{Perkins1995} and AFM \cite{Rief1997}. While the statistical mechanics of unconstrained DNA under tension is theoretically well understood in the framework of the
Worm Like Chain model \cite{Bustamante1994,Vologodskii1994,Marko1995,Odijk1995,Bouchiat1999} the presence of topological constraints
like supercoiling \cite{Bouchiat1998,Bouchiat2000,Moroz1997,Moroz1998} or geometrical constraints like
protein induced kinks and bends \cite{Bruinsma1999,Yan2003,Metzler2002,Kulic2004,Kulic2005a} renders
analytical results more difficult. 

Instead of studying the elastic properties of biopolymers under stretching, mechanical properties can also be studied by using compression, as long as the chains are smaller than the peersistence length. This has been for example used in experiments targeted to measure the force-velocity relation of microtubule growth \cite{Janson2004} and in determining the force produced by actin filaments \cite{Kovar2004}.

\begin{figure}[t] 
\centering
\includegraphics[scale=0.5]{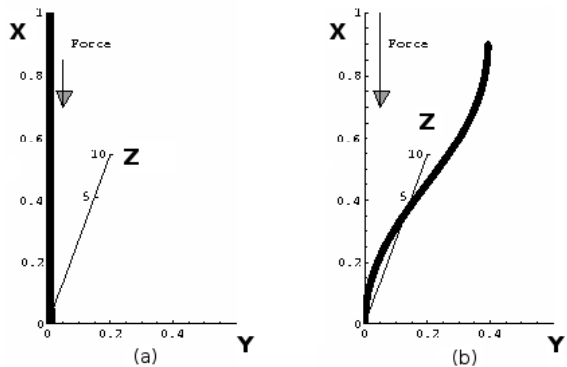}
\caption{Force below (a) and above (b) the Euler transition.}\label{fig:configuration}
\end{figure}

With the exception of the work of Odijk \cite{Odijk1998} who considers a semi-classical evaluation of the partition function in the linear regime (Fig. \ref{fig:configuration} (a)), i.e., below the buckling transition, no calculations have been done that consider the non-linear regime of  external forces above the critical force (Fig. \ref{fig:configuration} (b)). Furthermore these calculations are only valid well below the transition, although it is the behavior close to the transition on which the force calculations are based. In this paper we study thermal fluctuations up to the transition in order to evaluate the scaling of the point of buckling with increasing length.

A computer simulation for 2 and 3 dimensional configurations shows that the thermal fluctuations decrease the extension in the buckled state of the polymer in 3 dimensions, but increase it in 2 dimensions (Fig. \ref{fig:simulation}). In this paper we show analytically how this happens, by doing a harmonic perturbation calculation around the buckled state.

As a final note we mention that recently the properties of DNA, like its stiffness and its sequence-specific pairing have been exploited to build different kinds of nanostructures \cite{Seeman2003}. In particular a DNA tetrahedron which has been already synthesized could be the building block of extended nanostructures \cite{Goodman}. Our calculations can be used to estimate the forces these structures can withstand. 

The paper is organized as follows: we start by describing the geometry and the model used in chapter II, briefly treating its classical elastica solutions in chapter III. The main body of the paper consists of a semi-classical calculation of the force extension behavior for a WLC with finite length and persistence length below and above the Euler transition in chapter IV. We extend these calculations to quartic order below the transition in chapter V, in order to analyze the change in buckling transition caused by thermal fluctuations. In chapter VI we compare our calculations with simulations. In the concluding chapter VII we discuss our results in the light of several recent experiments with stiff biopolymers.

\bigskip

\section{The Partition sum of a worm like chain under compression}

We model a stiff polymer as a worm like chain without a twist
degree of freedom. In this case, the polymer configuration is completely characterized by
specifying the unit vector $\mathbf{t}(s)$ along the chain, where $s$ is the contour length with $0<s<L$, with $L$ being the chain length. When the chain is submitted to a compressive force $\fo$ the total energy is 
\begin{equation}
E\left[  \tv \right]  =\int\nolimits_{0}^{L}\left[  \frac{A}{2}\left(
\frac{d\tv}{ds}\right)  ^{2}+\mathbf{\fo.t}\right]  ds
\end{equation}
It is custom to write the bending stiffness as $A=l_{P}(T)k_{B}T$ where
$l_{P}(T)$ is the orientational persistence length and $k_{B}T=1/\beta$ the thermal
energy; for example for DNA at room temperature $l_{P}\approx 50nm$ \cite{Hagerman}.
All the statistical properties of interest can be deduced from the partition function which is a non-trivial quantity to evaluate because of the local constraint $\tv^2(s)=1$ that assures the inextensibility of the chain
\begin{equation}
Z=\int\delta^{(3)}\left(  \mathbf{t}^{2}-1\right)  \mathcal{D}%
^{3}\left[ \tv\right]  e^{-\beta E\left[  \tv\right]  }\label{Q}%
\end{equation}
This partition function is nothing but the Euclidean path integral of a quantum
particle with mass $A$, moving on a unit sphere under the influence of an
external constant force. We are interested in the thermal fluctuations around a ``classical path''. These are easiest to find in polar coordinates. We will fix the force along the $x$-axis (see Fig. \ref{fig:configuration} (a),(b)) to avoid the chart singularity at the poles. For notational convenience we will choose the polar angle $\vartheta \in [-\pi/2,\pi/2]$ such that the uncompressed chain has the coordinates $(\vartheta(s),\varphi(s))=(0,0)$. In these coordinates the energy has the form:
\begin{equation}\label{eq:energypolar}
E\left[\vartheta(s),\varphi(s)\right]=\int_0^L\left\lbrace \dfrac{A}{2}\left(\cos^2\vartheta(s)\overset{.}{\varphi}^2(s)+\overset{.}{\vartheta}^2(s)\right)+\fo\cos\vartheta(s)\cos\varphi(s)\right\rbrace ds
\end{equation}
We can rewrite the energy in dimensionless variables as:
\begin{align}\label{eq:energydimlesspolar}
\En\left[\theta(s),\phi(s)\right] &:= \dfrac{E\left[\theta(s),\phi(s)\right]}{k_BT}=\dfrac{1}{h}\int_0^{1}\left\lbrace \dfrac{1}{2}\left(\cos^2\theta(t)\overset{.}{\phi}^2(t)+\overset{.}{\theta}^2(t)\right)+G^2\cos\theta(t)\cos\phi(t)\right\rbrace dt\\
\phi(t)&:=\varphi(tL) \text{ , }\theta(t):=\vartheta(tL) \nonumber
\end{align}
where we have introduced the fluctuation parameter
$$h:= \dfrac{L}{l_p}$$
and the coupling strength
$$G:=L\sqrt{\dfrac{\fo}{A}}$$
The square root in the last expression is in fact the reciprocal of the deflection length \cite{Odijk1986} of the chain.
We are interested in the small fluctuation regime and use $h$ as an expansion parameter. The classical path will be the dominant path for $h\rightarrow0$, i.e., in the rod limit, and thermal fluctuations are taken into account by expanding the partition function in fluctuations around this classical path.
The partition function will now be a path integral in curvilinear coordinates \cite{Kleinert}:
\begin{equation} \label{eq:partition}
Z=\int\mathcal{D}^{2}\left[ \theta,\phi \right]  \sqrt{g\left(
\theta \right)  }e^{- \En\left[\theta(t),\phi(t)\right]  }
\end{equation}
The determinant of the metric in these coordinates is given by $g\left(\theta\right)  = \cos^2\theta$. The square root of this determinant, as present in the path integral measure, formally takes care of the coordinate independence (chart independence) of the measure. It can be understood in a time sliced version, although not without subtleties \cite{Kleinert}. This measure term can also be formally exponentiated resulting in an extra energy term:
\begin{equation}
Z=\int\mathcal{D}^{2}\left[ \theta,\phi \right] e^{- \En\left[\theta(t),\phi(t)\right]  -\En_m\left[\theta(t),\phi(t)\right]}
\end{equation}
with the measure energy term:
\begin{equation}\label{eq:measure-energy}
\En_m\left[\theta(t)\right]= -\delta(0)\int_0^1 dt \log\cos\theta(t) 
\end{equation}
The delta function in front of the integral should be understood as being finite using some regularization scheme.
The classical solutions are obtained through the Euler-Lagrange equations in the next section. We will then proceed by incorporating small fluctuations around these classical solutions in section (\ref{sec:semclass}).

As we will see there are values of the coupling strength where several classical solutions exist with comparable Boltzmann weight. These give rise to a bifurcation point in the groundstate. 
Also since the potential term in (\ref{eq:energydimlesspolar}) is positive, there are values of $G$ for which the actual groundstate breaks the rotational symmetry around the direction of the applied force. The associated goldstone modes can be excluded by explicitly fixing a direction.

\section{Euler buckling}

In this paper we consider as in Ref. \cite{Odijk1998} a molecule that has its two ends clamped at fixed orientations $\phi(0)=\phi(1)=\theta(0)=\theta(1)=0$, while the ends can freely move in the plane perpendicular to the force. In the zero fluctuation parameter limit the partition function gets only contributions from the classical paths, that minimize the energy. The Euler-Lagrange equations are:
\begin{align}\label{eq:eulerlagrange}
\overset{..}\theta(t)=-\cos\theta(t)\sin\theta(t)\overset{.}\phi^2(t)-G^2\sin\theta(t)\cos\phi(t)\\
\dfrac{d}{dt}\left(\cos^2\theta(t)\overset{.}\phi(t)\right)= -G^2\cos\theta(t)\sin\phi(t)
\end{align}
These equations can be integrated resulting in two classes of solutions: 
the straight rod solution
\begin{align}
\theta(t) &= 0& \phi(t)&=0 
\end{align}
and the buckling solutions that read by choosing $\theta(t)=0$
\begin{align}
 \df{\phi}^2(t)&=2G^2(\cos\phi(t)-1+2m), m \in [0.1) \Rightarrow \label{eq:buckdt} \\
 \phi(t)&=2\arcsin(\sqrt{m}\sn(tG|m)) \Rightarrow \label{eq:class}\\
 \cos\phi(t)& =1-2m\sn^{2}\left(tG |  m\right)\label{eq:buckling}
\end{align}
Here $\sn()$ is an elliptic Jacobi function \cite{Abramowitz}. Solutions with $m>1$ are solutions containing loops. They have a higher energy in our case. Using the periodicity properties of $\sn$ we find for buckling solutions with the boundary condition $\phi(1)=0$ the following relation between $m$ and $\fo$
\begin{align}\label{eq:force}
G= L\sqrt{\dfrac{\fo}{A}}&=2n\K(m)&  n\in \mathbb{Z}
\end{align}
Here $\K(m)$ is the complete elliptic integral of the first kind. We will label the solutions with $n$, $\phi_0$ corresponding to the straight rod. Since $\K(m)$ is a monotonously increasing function of $m$ we find a smallest force that permits a given buckling solution \cite{Love}:
\begin{equation}\label{eq:critical}
\fo_c= G^2\dfrac{A}{L^2} =n^2\frac{4\K(0)^2}{L^2}A= n^2\dfrac{\pi^2}{L^2}A
\end{equation}
It is straightforward to calculate the end-to-end distance along the $z$-axis by integrating the solution along the chain. The result is:
\begin{align}\label{eq:extension}
X&= L\int_0^1dt\cos\phi(t) \nonumber \\
&=L\left(\dfrac{2\E(m)}{\K(m)}-1 \right)
\end{align}
The value of the extension becomes negative under large enough compression. In practice we will be interested in the region where the force is small enough that the WLC model is still reasonable. It is easy to see from the buckling solution (\ref{eq:buckling}) that the compressed chain will be a monotone curve as long as $m < 1/2$. For higher values of $m$ the chain forms an $s$-shaped curve.

The energy of the buckling solution is found from (\ref{eq:energypolar}) to be:
\begin{align}\label{eq:buckle-energy}
\En_n(m)&= \dfrac{G^2} {h} \left(2\dfrac{X}{L}+2m-1\right) \nonumber \\
&=\dfrac{4n^2\K^2(m)}{h}\left(4\dfrac{\E(m)}{\K(m)}+2m - 3\right)
\end{align}
with $m$ depending on the force, $\fo$, and $n$ through (\ref{eq:force}). $\E(m)$ is the complete elliptic integral of the second kind.

When comparing the energy of the buckled state with the straight rod configuration we notice that the buckled state is always energetically favorable once it is allowed by (\ref{eq:critical}). This transition from straight rod to the buckled state is referred to as the Euler transition. When no other constraints are imposed on the solutions the first buckling solution, $n=1$, will be the favorable solution under compression once the first critical value for the force has been reached \cite{Love}. When the end of the chain is constrained to be fixed in the origin of the $YZ$-plane, making both ends fixed on the $z$-axis, it is the one loop solution that, when there are no constraints on the rotation of the chain around its axes, is the favorable solution. We will for the rest of this article restrict ourself to the unconstrained case.

\section{Semiclassical buckling }\label{sec:semclass}

For finite values of the fluctuation parameter thermal fluctuations must be taken into account in the evaluation of the partition function. We will write the coordinates as:
\begin{align}
\theta(t)=&\theta_n(t)+\de \theta(t)=\de \theta(t) & \phi(t) &= \phi_n(t) +\de \phi(t)
\end{align}
Here the index $n\in \mathbb{Z}$ differentiates between the classical solutions (\ref{eq:class}-\ref{eq:force}), the straight solution corresponding to $n=0$.
By plugging these relations into the expression for the total energy (including the measure term) we find order by order:
\begin{align}\label{eq:perturb}
\En&[\theta(t),\phi(t)]+\En_m[\theta(t)]= \nonumber \\
&\dfrac{1}{h}\int_0^1dt\left\lbrace \dfrac{1}{2}\df{\phi}_n^2+G^2\cos\phi_n\right\rbrace \nonumber \\
+& \dfrac{1}{h}\int_0^1dt\left\lbrace \df{\phi}_n\de\df{\phi} - G^2\sin\phi_n\de\phi\right\rbrace \nonumber \\
+& \dfrac{1}{h}\int_0^1dt\left\lbrace \dfrac{1}{2}(\de\df{\phi})^2-\dfrac{1}{2}G^2\cos\phi_n(\de\phi)^2 + \dfrac{1}{2}(\de\df{\theta})^2 - \dfrac{1}{2}(G^2\cos\phi_n+\df{\phi}_n^2)(\de\theta)^2\right\rbrace \nonumber \\
 +&\cdots
\end{align}
The first term is just the energy as given by (\ref{eq:buckle-energy}) for the buckle solutions. The second term is zero when we look at chains with fixed boundary conditions (Dirichlet boundary conditions). The third term represents the lowest order that accounts for thermal fluctuations and is in the focus of our attention. Note that the measure term will only show up in the quartic order fluctuations (of order $h$), since for the Gaussian distribution the fluctuations are of order $\sqrt{h}$

\subsection{Harmonic fluctuations below the Euler transition}
We first consider the regime below the critical force $\fo_c$, $G <G_c=\pi$, where the classical solution is the straight rod. The partition function to lowest order around this ground state has the simple form:
\begin{equation}
Z=\exp(-G^2/h )\int \mathcal{D}[\de\theta,\de\phi] \exp\left\lbrace -\dfrac{1}{h}\int_0^1dt\left(  \dfrac{1}{2}(\de\df{\theta})^2-\dfrac{1}{2}G^2(\de\theta)^2\right)\right\rbrace   \exp\left\lbrace -\dfrac{1}{h}\int_0^1dt \left(\dfrac{1}{2}(\de\df{\phi})^2-\dfrac{1}{2}G^2(\de\phi)^2\right)\right\rbrace 
\end{equation}
The resulting path integral is the product of the partition sums, in Euclidean time, of 2 independent harmonic oscillators with a frequency squared of $-G^2$.
When we consider first the azimuth, $\phi$, contribution it is in fact a harmonic oscillator on the circle where angles that differ a full period are equivalent. The pathintegral in that case can be expressed as a sum over the harmonic oscillator on the real line by summing over all equivalent end points (see e.g. \cite{Kleinert} chapter $6$):
\begin{align}
Z_{\text{circle}} (\phi(0)=0,\phi(1)=0)&=\sum_{n=-\infty}^{+\infty} Z_{\text{line}}(\phi(0)=0,\phi(1)=2\pi n)\\
&= \dfrac{1}{\sqrt{2\pi h}}\sqrt{ \dfrac{G}{\sin G} } \vartheta_3(0,e^{-2\pi^2G\cot G/h})
\end{align}
The elliptic theta function, $\vartheta_3(0,q)$ \cite{Gradshteyn} diverges for $G=\pi/2$, only half the critical force, which seems to be odd at first sight. The reason behind this is that for $G=\pi/2$ all equivalent paths have the same weight, there is no cost in increasing the winding number. As a first correction we note that higher order corrections considerably temper the potential abyss for larger fluctuations in which case we can neglect the contributions from the winding by taking the domain of $\phi$ to be the real line. This results in an improved estimate for the partition sum:
\begin{align}
Z_{\phi} = \dfrac{1}{\sqrt{2\pi h}}\sqrt{ \dfrac{G}{\sin G} }
\end{align}

The same kind of reasoning holds for the polar angle. Here we do not have winding, but formally an oscillator in a box. Since we again assume the fluctuations to be small it is possible to extend the domain to the real axis. Although we have the equivalence $(\theta,\phi) \sim (\pi -\theta, \phi + \pi)$ again the results do not hold for larger fluctuations that have a weight that does differ substantially from zero. So by taking the polar angle also covering the real line we are only overcounting configurations that do not contribute to the path integral. The final result is then:
\begin{equation}\label{eq:hampart}
Z=\exp(-G^2/h )  \dfrac{1}{2\pi h} \dfrac{G}{\sin G} 
\end{equation}
This partition sum diverges at the caustics, $G=\pi$. Note that this is exactly the critical point for Euler buckling. Here it is caused by the harmonic potential being just strong enough to cancel the kinetic term (i.e. the bending energy), making large fluctuations favorable and thus invalidating the harmonic approximation. Unlike the $\vartheta_3$-function divergence here we can not just dismiss these larger fluctuations, since they do not come from a topological disconnected region in configuration space and as such are indeed an indication that the groundstate is suffering from an instability.

The force extension behavior is readily obtained, as an approximation, from the partition function:
\begin{align}\label{eq:forceextrod}
X(\fo)&=-\dfrac{1}{\beta Z}\dfrac{\partial Z}{\partial \fo} \nonumber \\
&= L \left( 1- \dfrac{h}{2G^2}(1-G\cot G) \right)
\end{align}
This expression diverges again at the Euler transition. Since we have approximated the total extension $X=L\int dt\cos\theta \cos \phi$ to quadratic order in the fluctuations around the classical solution, the deviation of above expression for the extension from the straight rod actually gives, up to a factor $L$, the variance of the fluctuations averaged over the chain. When this variance is large not only the harmonic approximation to the partition sum breaks but the force extension approximation breaks down as well. From these considerations we expect the above force-extension relation to hold as long as $\pi-G \gg h/2\pi$. From this observation one is tempted to conclude that the rod will start to buckle at a force shifted downwards from the Euler transition force following a scaling law for small $h$ of :
\begin{align}
\fo_c\sim \fo_c^{(0)}(1-Ch)
\end{align}
with $C$ a constant of order $1$. This is a well known result from Ref. \cite{Odijk1998}. We will have to adjust this picture when taking higher order terms into account, as we will see in section \ref{ch:quartic}, because the linear scaling tells us only something about the validity of the quadratic approximation.

For small forces, $G\ll 1$, we find from (\ref{eq:forceextrod}) for the extension of the chain:
\begin{align}\label{eq:lowforce}
X(\fo) \cong L\left(1-\dfrac{h}{6}(1+\dfrac{G^2}{15})\right) 
\end{align}
For $G=0$ this is the extension of the chain shortened by thermal fluctuations alone.

\subsection{Harmonic fluctuations above the Euler transition}

The harmonic correction to the classical solution has again the form of an harmonic oscillator, but now with a ``time'' dependent oscillator frequency. The azimuth and polar part of the fluctuation factor again decouple:
\begin{equation}
Z=e^{-\En_1(m)}F_{\phi}F_{\theta}
\end{equation}
The classical solution is given by (\ref{eq:buckle-energy}). In principle we should sum over all classical buckling solutions that are allowed at a given force. The energy difference is nonetheless big enough that we can neglect the contribution of  higher buckled configurations.

The azimuth contribution has the form (after partial integration):
\begin{equation}
F_{\phi}=\int\mathcal{D}[\de\phi]\exp\left(-\dfrac{1}{2h}\int_0^1dt\de\phi\mathbf{\hat{T}}_{\phi}\de\phi\right)
\end{equation}
with the harmonic fluctuation operator given by
\begin{equation}
\mathbf{\hat{T}}_{\phi}=-\dfrac{d^2}{dt^2}-G^2\cos\phi_1(t)
\end{equation}
where $\phi_1(t)$ is the classical $n=1$ buckling solution.
The fluctuation factor can be written using Gaussian integration in terms of functional determinants as:
\begin{equation}
F_{\phi}=\dfrac{1}{\sqrt{2\pi h}}\left[\dfrac{\det(-\dfrac{d^2}{dt^2})} {\det(\mathbf{\hat{T}}_{\phi})}\right]^{1/2}
\end{equation}
The determinant of the fluctuation operator can be calculated using the Gelfand-Yaglom formula \cite{Kleinert}. 
To do so we have to find a solution $D_{\phi}(t)$ of the differential equation  
\begin{equation}
\mathbf{\hat{T}}_{\phi}D_{\phi}(t)=0
\end{equation}
with boundary conditions $D_{\phi}(0)=0$ and $\df{D}_{\phi}(0)=1$. The determinant $\det(\mathbf{\hat{T}}_{\phi})$ is then given by $D_{\phi}(1)$. Changing variables to $x=Gt$ the differential equation has the form of a Lam\'e equation \cite{Arscott} (the Laplacian in ellipsoidal coordinates):
\begin{equation}
\dfrac{d^2y(x)}{dx^2} +\{1-2m\sn^2(x|m)\}y(x)=0
\end{equation}
With the given coefficients there exists one double periodic solution (also called Lam\'e polynomial) given by a Jacobi elliptic function
\begin{equation}
y(t)=\cn(Gt)
\end{equation}
This solution has not the right boundary conditions, but using D'Alemberts construction \cite{Kleinert}, that gives another independent solution, we can construct the solution with the right boundary conditions:
\begin{align}
D_{\phi}(t)&=y(t)y(0)\int_0^t\dfrac{dt'}{y^2(t')} \nonumber \\
&=\dfrac{\sn(Gt|m)\dn(Gt|m)-\E(Gt|m)\cn(Gt|m)}{G(1-m)}+ t\cn(Gt|m)
\end{align}
Here we adhere to the notation for the Elliptic Integral of the second kind as used in Abramowitz \cite{Abramowitz}, see also appendix (\ref{app:elliptic}). The function $\dn$ is the last Jacobi elliptic function we need.

With this solution to the Lam\'e equation we find the fluctuation determinant as:
\begin{align}
D_{\phi}:=\dfrac{\det \left(\mathbf{\hat{T}}_{\phi}\right)}{\det(-\dfrac{d^2}{dt^2})} &= D_{\phi}(1) \nonumber \\
&= \dfrac{\sn(G|m)\dn(G|m)-\E(G|m)\cn(G|m)}{G(1-m)}+ \cn(G|m) 
\end{align}
Now we can make use of the relation $G=2\K(m)$ (\ref{eq:force}) to simplify this result to
\begin{equation}
D_{\phi} = \dfrac{\E(m)-(1-m)\K(m)} {(1-m)\K(m)}
\end{equation}
from which we obtain the fluctuation factor:
\begin{equation}\label{eq:phifluc}
F_{\phi}=\sqrt{\dfrac{(1-m)\K(m)}{2\pi h (\E(m)-(1-m)\K(m))}}
\end{equation}
Since for small $m$, $\E(m)- (1-m)\K(m) \sim m\pi/4$, the fluctuation factor diverges at the Euler transition. This is not too surprising since in the 2 dimensional configuration, there are with forces close to the buckling transition three classical solutions with comparable energies with only small barriers in between, allowing larger thermal fluctuations than admissible for a harmonic approximation. Would we forbid out-of-plane fluctuations the picture is that fluctuations would grow with increasing force just below the Euler transition. Just above the Euler transition the chain will fluctuate between the two possible buckled configurations, analogous to quantum tunneling. Finally the buckling will stabilize with increasing force to one of the two configurations. 

We now come to the out-of-plane fluctuations. The fluctuation determinants can again be calculated using the Gelfand-Yaglom method.
We are now looking for a solution of (with $x=Gt$):
\begin{equation}
\dfrac{d^2y(x)}{dx^2} +\{1+4m-6m\sn^2(x|m)\}y(x)=0
\end{equation}
This happens to be again a Lam\'e equation with the right coefficients to have a double periodic Lam\'e polynomial as solution:
\begin{equation}\label{eq:zeromode}
y_0(x)=\sn(x)\dn(x)
\end{equation}
Since $y_0(0)=0$ we immediately find for the fluctuation determinant:
\begin{equation}
D_{\theta}:=\dfrac{\det \left(\mathbf{\hat{T}}_{\theta}\right)}{\det(-\dfrac{d^2}{dt^2})} = \dfrac{\sn(G)\dn(G)}{G}\equiv 0
\end{equation}
and the partition sum diverges. This is caused by the global rotations around the force direction connecting a continuum of groundstates. 
The buckling solution (\ref{eq:buckling}) was chosen to be lying in the $xy$-plane. Since the energy (as well as the pathintegral measure) is invariant under rotations around the x-axis we have a continuum of buckling solutions. We can make use of this symmetry by integrating only over paths where the angle $\theta$ averages along the chain to zero and then integrating separately over the rotation around the $x$-axis. This can be done in a consistent way using the Faddeev-Popov (FP) method \cite{Peskin} developed to fix internal symmetries in quantum field theory.
A clockwise rotation of the chain by an angle $\gamma$ around the $x$-axis changes the coordinates on the sphere to:
\begin{align}\label{eq:rotations}
\cos\theta\sin\phi &\rightarrow \cos(\theta_{\gamma})\sin(\phi_{\gamma})=\cos\theta\sin\phi\cos\gamma + \sin\theta\sin\gamma \nonumber\\
\sin\theta &\rightarrow \sin(\theta_{\gamma})=-\cos\theta\sin\phi\sin\gamma + \sin\theta\cos\gamma
\end{align}
Now we want to fix the average of the $\theta$ angle, $\bar{\theta}:=\int_0^1dt\theta(t)$, to zero. We define the FP ``determinant'' through:
\begin{equation}\label{eq:FPdef}
\Delta_{FP}[\theta,\phi] \int_0^{2\pi} d\gamma \de(\bar{\theta}_{\gamma}) = 1
\end{equation}
where the argument of the delta function is the average angle of the by $\gamma$ rotated chain. Inserting ``$1$'' into the partition sum (\ref{eq:partition}) results in:
\begin{align}
Z&=\int_0^{2\pi} d\gamma\int\mathcal{D}^{2}\left[ \theta,\phi \right]  \sqrt{g\left(
\theta \right)} \Delta_{FP}  \de(\bar{\theta}_{\gamma}) e^{- \En\left[\theta(t),\phi(t)\right]  } \nonumber \\
&= 2\pi \int\mathcal{D}^{2}\left[ \theta,\phi \right]  \sqrt{g\left(\theta \right)} \Delta_{FP}  \de(\bar{\theta}) e^{- \En\left[\theta(t),\phi(t)\right]  }
\end{align}
In the last step we have first performed a trivial change of variable of integration and then made use of the invariance under rotation of the energy and of the pathintegral measure. In fact just the invariance of the combination of the measure and the Boltzmann factor would have been enough.
The FP determinant can be found from the definition (\ref{eq:FPdef}) :
\begin{align}
\Delta_{FP}[\theta,\phi]&=\left(\int_0^{2\pi} d\gamma \de(\bar{\theta}_{\gamma})\right)^{-1} \nonumber \\
&=\left|\int_0^1 dt \sin\phi(t)\right|
\end{align}
Since we are interested in small thermal fluctuations around the classical solution we can assume the fluctuations to be such that $\int dt\sin\phi(t) >0$ for all relevant paths. This apparently does not hold anymore close to the bifurcation point. 
Defining $Z_0$ to be the partition sum without the FP term, but including the angle fixing delta function, the lowest order contribution of the FP term to the partition sum is:
\begin{align}\label{eq:fadpoppartition}
Z &= \int_0^1dt\left\langle \sin\phi(t) \right\rangle Z_0 \nonumber \\
&\cong \int_0^1dt\sin\phi_1(t) Z_0 \nonumber\\
&= 2\sqrt{m}\int_0^1 y_0(Gt) Z_0 =: Z_{FP}Z_0
\end{align}
The last step follows from the definition of $\phi_1(t)$ (\ref{eq:buckling}).
We now fix the global polar angle in the polar fluctuation factor:
\begin{equation}
F_{\theta}:=2\pi Z_{FP}\int\mathcal{D}[\de\theta]\de(\de\bar{\theta})\exp\left(-\dfrac{1}{2h}\int_0^1dt\de\theta\mathbf{\hat{T}}_{\theta}\de\theta \right)
\end{equation}
To see how this procedure formally gets rid of the divergence we note first that the fluctuation operator, as defined on the square integrable functions on $[0,1]$ that are zero on the boundary, is symmetric and so we can find a real orthonormal basis $\left\lbrace \tilde{y}_n \right\rbrace$ that diagonalizes the operator. Using this basis we write $\de\theta(t)=\sum_{n=0}^{\infty} x_n\tilde{y}_n(Gt)$. The normalized zero mode eigenfunction is given by $\tilde{y}_0(Gt)=(\int dt y_0^2(Gt))^{-1/2}y_0(Gt)$ and the eigenvalues are written as $\lambda_n$,e.g. $\lambda_0=0$. We now integrate separately over the zero mode:
\begin{align}
F_{\theta}&=\dfrac{2\pi Z_{FP}}{\sqrt{2\pi h}}\left(\prod_{n\geq 0}\int \dfrac{dx_n}{\sqrt{2\pi h}}\right)\int\ \dfrac{dx_0}{\sqrt{2\pi h}}\quad\de\Big(\sum_n x_n\int_0^1 \tilde{y}_n(t)\Big)\exp\left(-\dfrac{1}{2h}\sum_{n\geq 1}x_n^2 \lambda_n \right)\nonumber \\
&=\dfrac{\sqrt{2 \pi} Z_{FP}}{\sqrt{h} \left|\int_0^1 dt \tilde{y}_0(Gt)\right|}\dfrac{1}{\sqrt{2\pi h}} \left(\prod_{n\geq 0}\int \dfrac{dx_n}{\sqrt{2\pi h}}\right)\exp\left(-\dfrac{1}{2h}\sum_{n\geq 1}x_n^2 \lambda_n \right) \nonumber \\
&= \dfrac{2\sqrt{2 m \pi \int_0^1dt y^2_0(Gt)}}{\sqrt{h} }\lim_{\epsilon \rightarrow 0}\sqrt{\dfrac{\lambda_0^{\epsilon}}{2\pi h D_{\theta}^{\epsilon}}} \label{eq:thetafluc}
\end{align}
In the last step we regularized the determinant by adding a small linear term:
\begin{align}
\mathbf{\hat{T}}_{\theta}^{\epsilon} := \mathbf{\hat{T}}_{\theta} + \epsilon \hat{1}
\end{align}
in effect shifting all eigenvalues $\lambda_n$ by $\epsilon$ to the new values $\lambda_n^{\epsilon}=\lambda_n +\epsilon$. The resulting determinant is then, in first order in $\epsilon$, $\epsilon$ times the determinant of the reduced operator defined on the orthogonal complement of the zero mode eigenvector, since all other linear terms contain the zero mode eigenvalue. The resulting homogeneous differential equation has been solved for a similar case in \cite{Kulic2005}. The somewhat technical calculation is done in an appendix \ref{app:genlam}. The resulting determinant is:
\begin{equation} \label{eq:thetadet}
D_{\theta}^{\epsilon}= \dfrac{\epsilon}{\K(m)3m}((1-m)\K(m) - (1-2m)\E(m))
\end{equation}
Finally the integral over the zero mode squared is given by:
\begin{align}\label{eq:jacobian}
\left(\int_0^1 dt y_0^2(t)\right)^{1/2} &= \left(\dfrac{2}{3}[(1-m)\K(m)-(1-2m)\E(m)]\right)^{1/2}
\end{align}
Combining (\ref{eq:phifluc}), (\ref{eq:thetafluc}), (\ref{eq:thetadet}) and (\ref{eq:jacobian}) we find for the partition sum:
\begin{equation}\label{eq:totpart}
Z=e^{-\En_1}\dfrac{2m\K(m)}{\pi}\sqrt{\dfrac{(1-m)} { h^3 [\E(m)-(1-m)\K(m)]} }
\end{equation}
It is noteworthy that the partition sum does not diverge at the Euler transition, but goes to zero. By approximating the Faddeev-Popov determinant by its classical value we are in fact underestimating the amount of configurations the closer we come to the bifurcation point. 
The force extension corrections to the classical force extension curve $X_0(\fo)$, equation (\ref{eq:extension}), defined as $X=X_0+X_{\phi}+X_{\theta}$, with the subscript labeling the fluctuation part that causes the extension change, are given by:
\begin{figure} 
\centering
\includegraphics{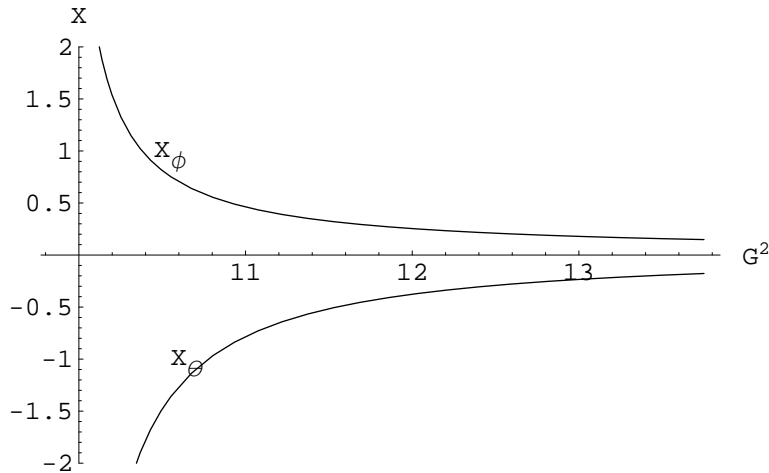}
\caption{Relative extension shift from Eqs. (\ref{eq:angleextension}) with $h=L=1$}\label{fig:extensionshift}
\end{figure}
\begin{align}\label{eq:angleextension}
X_{\phi}&\cong -\dfrac{1}{\beta F_{\phi}}\dfrac{\partial F_{\phi}}{\partial \fo}= hL\dfrac{m(1-m)}{16\K(m)(\E(m)-(1-m)\K(m))}\left(\dfrac{\K(m)}{\E(m)-(1-m)\K(m)}-\dfrac{\E(m)-(1+m)\K(m)}{m(1-m)\K(m)}\right)\nonumber\\
X_{\theta}&\cong -\dfrac{1}{\beta F_{\theta}}\dfrac{\partial F_{\theta}}{\partial \fo}= -hL\dfrac{(\E(m)+3(1-m)\K(m)}{16\K^2(m)(\E(m)-(1-m)\K(m))}
\end{align}
These formula are not too illuminating. Plotting the two corrections (Fig. \ref{fig:extensionshift}) reveals that the corrections to the extension caused by thermal fluctuations have an opposite sign. The out-of-plane fluctuations make the chain slightly shorter than the classical solution, as is to be expected. The in-plane fluctuations have the opposite effect. This can be understood as the extension change by fluctuations in the straight rod direction to be stronger than fluctuations away from the rod solution.

The total extension again diverges when approaching the bifurcation point, both for the $X_{\phi}$ and $X_{\theta}$ part seperately. For the azimuth part the reason behind this is the same as in the straight rod case: near the bifurcation point fluctuations increase because the two classical solutions, of positive and negative angle, are close to each other and as such a quadratic approximation to the force term is not enough. For the polar angle this is not the case since we integrated out the fluctuations to equivalent states, but there the FP term (\ref{eq:fadpoppartition}) is underestimated: as long as the deviation of the expectation value of the end point of the chain (proportional to the FP term) from the straight rod is larger than its fluctuations we can expect that our results hold. Close to the bifurcation point however, we are not allowed to drop the absolute value sign by going to Eq. (\ref{eq:fadpoppartition}) and find a lower bound of the FP term in the order of the standard deviation of the end point.

For small $m$, approaching the bifurcation point, we find from (\ref{eq:angleextension}):
\begin{align}\label{eq:extdifftrans}
X_{\phi} &= \dfrac{hL}{\pi^2 m}\left(1-\dfrac{m}{2} +\mathcal{O}(m^2)\right) \nonumber \\
X_{\theta}& = -\dfrac{hL}{\pi^2 m}\left(2-\dfrac{5m}{2} +\mathcal{O}(m^2)\right)
\end{align}

Like below buckling the extension diverges because we make an approximation by taking the extension to be $-\frac{1}{\beta}\partial_{\fo}\log Z$. This is not exact when approximating the potential. For the same reasons as below buckling we can expect he results not to hold for large relative extension shifts.

\section{Quartic order}\label{ch:quartic}
Below buckling it is fairly simple to get a good estimate of the force extension curve up to the Euler transition by taking higher order fluctuations into account. Since it is the lowest mode that is responsible for the blowing up of the partition sum, approaching the transition, we can significantly improve the calculations by including the quartic term for this mode. Quartic terms containing other modes hardly improve upon this. In $2$ dimensions the corrected partition sum is:
\begin{align}
Z&=\dfrac{e^{-G^2/h}}{\sqrt{2\pi h}}\sqrt{\dfrac{G(\pi^2-G^2)}{\sin G}} \int_{-\infty}^{\infty}\dfrac{dx}{\sqrt{2\pi h}}\exp(-\dfrac{1}{2h}(x^2(\pi^2-G^2)+x^4\dfrac{G^2}{8})\nonumber \\
&=\dfrac{e^{-G^2/h}}{\sqrt{2\pi h}}\sqrt{\dfrac{G(\pi^2-G^2)}{\sin G}}\dfrac{\gamma}{2\tau\sqrt{\pi h}}e^{\frac{\gamma^4}{2\tau^2}} \K_{\frac{1}{4}}\left(\frac{\gamma^4}{2\tau^2}\right)
\end{align}
with
\begin{align}
 \tau&=\dfrac{G}{4\sqrt{h}}&  \gamma&=\dfrac{\sqrt{\pi^2- G^2}}{2\sqrt{h}}
\end{align}
From which we find for the force extension relation:
\begin{align}
X_{\gamma}&=L\left\lbrace 1-\dfrac{h}{2G}\left[\dfrac{\pi^2-3G^2}{2G(\pi^2-G^2)} -\frac{\cot G}{2}+\frac{1}{\gamma}\frac{d\gamma}{dG}-\frac{1}{\tau}\frac{d\tau}{dG} +(2-\dfrac{\K_{3/4}(\frac{\gamma^4}{2\tau^2})+\K_{5/4}(\frac{\gamma^4}{2\tau^2})} {\K_{1/4}(\frac{\gamma^4}{2\tau^2})})\frac{\gamma^4}{2\tau^2}\left(  \frac{2}{\gamma}\frac{d\gamma}{dG}-\frac{1}{\tau}\frac{d\tau}{dG}  \right)\right] \right\rbrace \label{eq:fextquad}
\end{align}

These solutions can be continued above the transition, but they start to deviate fast from the exact values. The more practical use of these calculations is to make an estimate of the forces a rod can endure, before it collapses.
Assuming $h\ll 1$, so that we are close to a buckling type of behavior, we can recognize two separate asymptotic regions of behavior, depending on the argument of the modified Bessel functions in (\ref{eq:fextquad}): 
\begin{itemize}
\item $\pi^2-G^2 \ll G\sqrt{h} $ we find as asymptotic behavior:
\begin{align}\label{eq:forcebuck}
X\simeq L\left(1-\frac{2\sqrt{2}}{\gam(1/4)^2}\sqrt{h}+(\frac{2}{\pi}+\frac{2\sqrt{2}\gam(-1/4)}{\gam(1/4)^3}+\mathcal{O}(\sqrt{h}))(\pi-G) +\mathcal{O}((\pi-G)^2)\right)
\end{align}
i.e. $X$ exhibits a finite negative slope. The decrease of the extension with increasing force is substantial. The polymer can be considered to buckle. 
\item $\pi^2-G^2 \gg G\sqrt{h} $ In this region the decrease of the extension is of order $h$, the force extension curve being almost flat. There is no buckling yet.
\end{itemize}
The crossover region and thus the region where the buckling transition is located, is where this argument is of order unity. It is of course not possible to pinpoint a precise transition value, but the scaling of the transition shift follows from these observations: the force where the instability appears is shifted by thermal fluctuations according to: 
\begin{align} \label{eq:transforce}
\fo_c \sim \fo_c^{(0)}(1-C\sqrt{h})
\end{align}
\begin{figure}[t] 
\centering
\includegraphics[scale=1.5]{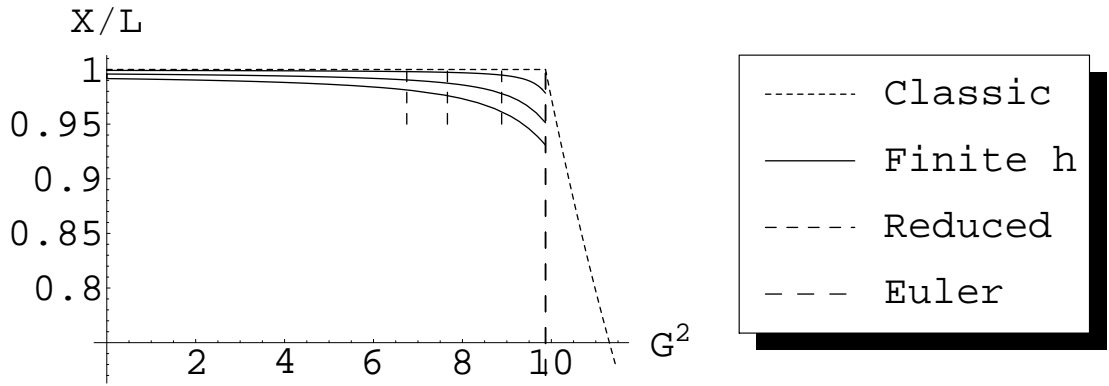}
\caption{Force extension in $2$-d for $h=0.01$, $0.05$ and $0.1$. The corresponding reduced transition forces from Eq. (\ref{eq:transforce}), with $C=1$, are shown by the $3$ short dashed vertical lines}\label{fig:reduce}
\end{figure}
With $C$ of order unity. The results for $3$ different values of $h$ are drawn in figure \ref{fig:reduce} together with the corresponding reduced transition forces using $C=1$.

We next consider the $3$d case. The contribution from the $\theta$ part alone is the same as for the $\phi$ part, which would result in a doubling of the difference from the straight rod. But now we also have a term mixing the two lowest modes. The fluctuation part of the partition sum is (apart from a constant):
\begin{align}\label{eq:4thzfl}
Z_{fl}&= \dfrac{G(\pi^2-G^2)}{\sin G} \int_{-\infty}^{\infty}dx \int_{-\infty}^{\infty}dz\exp(-\dfrac{1}{2h}(x^2\lambda_1+x^4\dfrac{G^2}{8}+z^2\lambda_1+z^4\dfrac{G^2}{8}+x^2z^2(\dfrac{3G^2}{4}-\dfrac{\pi^2}{2}))
\end{align}
As first approximation we can use the expectation value of the square of one of the modes, resulting in a modified $\gamma$ for the other mode given by:
\begin{align}\label{eq:3dparm}
\bar{\gamma} = \dfrac{1}{2\sqrt{h}}\sqrt{\pi^2-G^2+ \vev{z^2}\dfrac{(3G^2-2\pi^2)}{4}}
\end{align}
With:
\begin{align}
\vev{z^2}&= \left( \dfrac{\K_{3/4}(\frac{(\pi^2-G^2)^2}{2hG^2})+\K_{5/4}(\frac{(\pi^2-G^2)^2}{2hG^2})}{2\K_{1/4}(\frac{(\pi^2-G^2)^2}{2hG^2})}-1\right) \dfrac{2(\pi^2-G^2)}{G^2}-\dfrac{h}{\pi^2-G^2}
\end{align}
The resulting extension is then given by:
 \begin{align}\label{eq:average}
X=X_{\gamma}+X_{\bar{\gamma}}-L
\end{align}
This approximation slightly overestimates the contribution of the mixing term close to the transition, where the behavior is far from Gaussian. A better result can be obtained by treating the mixing term as a perturbation and expanding  \ref{eq:4thzfl}. The resulting series expansion one obtains is:
\begin{align}\label{eq:series}
Z_{fl}=\frac{G(\pi^2-G^2)}{\sin G}\sum_{n=0}^{\infty}\frac{1}{n!}\left(-\frac{3G^2-2\pi^2}{8h}\right)^n\left(\frac{\tau \gam(\frac{2n+1}{4})\Hyp(\frac{2n+1}{4},\frac{1}{2},\frac{\gamma^4}{\tau^2})-2\gamma^2\gam(\frac{2n+3}{4})\Hyp(\frac{2n+3}{4},\frac{3}{2},\frac{\gamma^4}{\tau^2})} {2\tau^{(2n+3)/2}}\right)^2
\end{align}
with $\Hyp(x,y,z)$ Kummer's function (confluent hypergeometric function). This series converges relatively fast just below the Euler transition and one can get a good approximation of the extension below the transition force. For practical purposes the first approximation is good enough to characterize the transition shift. It scales with increasing length in the same way as the $2d$ case.
We will next compare the predictions of the force-extension realtions, Eqs. (\ref{eq:forceextrod}), (\ref{eq:angleextension}),(\ref{eq:forcebuck}),(\ref{eq:average}) and (\ref{eq:series}) with simulations.

\section{Comparison with the simulation}
\begin{figure}[t] 
\centering
\includegraphics[scale=0.7]{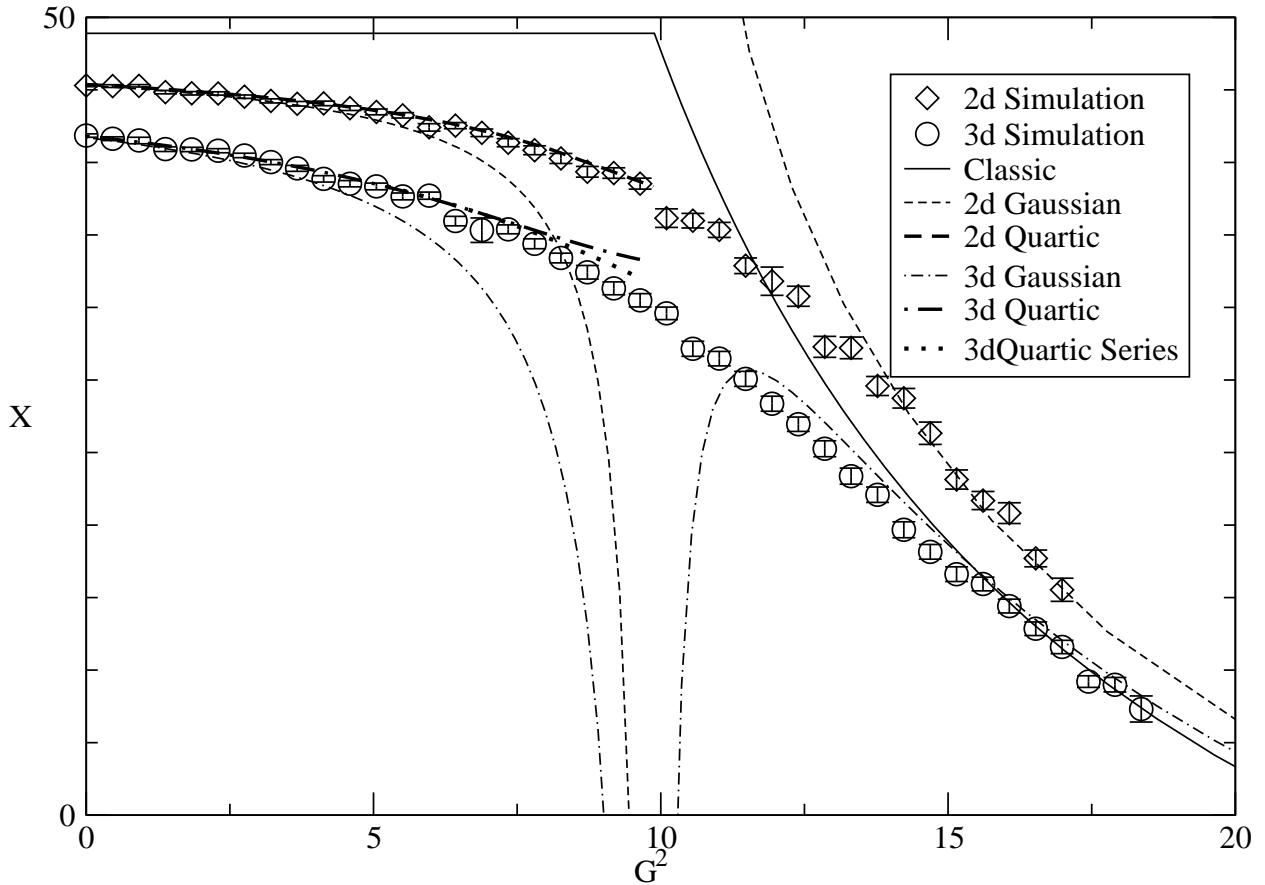}
\caption{Comparison of the analytical force-extension with simulations for $L=49$ and $h=0.8$. The unit of length is the bond length in the simulation.}\label{fig:simulation}
\end{figure}
The simulations were done with bond and bead models, consisting of 50 beads joined by either a bond consisting of a strongly repulsive Lennard-Jones potential and an attractive FENE potential \cite{kremer:5057,Byrd}, or with a stiff harmonic bond. The bending potential was implemented by a cosine angular energy term with a magnitude chosen so that the persistence length was comparable to the chain length ($L=49$ in terms of the bond length). The backbone stretching parameters were chosen in such a way that possible fluctuations of the bond length can be neglected compared to the bending fluctuations. The inextendible worm-like chain can then be expected to be a reasonable approximation to the simulated chain. The general features of the simulation did not depend on the type of bond chosen.

The simulations and theoretical calculations are plotted in Fig. \ref{fig:simulation}. The value of $h$ was taken rather high in order to have a more pronounced fluctuation contribution. The length scale is chosen such that the bond length in the simulation model is $1$. The $3$-d quartic curve was calculated using the modified quartic term. 

The semi-classical results are in good agreement with the simulation data in the region where a semi-classical approximation is expected to be valid. It is noteworthy that the increase in extension as predicted by the calculations is indeed the same as observed in the simulation. In $2d$ the quartic corrections below buckling show even for relatively large values of $h$, good agreement with simulations. In $3d$, using the simplified approach of modifying the quartic interaction to account for the mode mixing (\ref{eq:average}) the reliability of the calculations close to the Euler transition decreases, although the qualitative behavior seems to be good enough for practical purposes. Better results one gets using a perturbation expansion (\ref{eq:series}). The $3$-d quartic series curve was calculated using this expansion with the first $20$ terms. Note though that this last calculation was stopped slightly below the transition force, since it does not converge at the transition. 
 
The effect of the bond length not being fixed is indeed small enough compared to the thermal fluctuations. The errorbars are caused by the finite number of simulation rounds.

\section{Discussion}
The parameter that determines whether a buckling transition is present is the ratio $h$ of length and persistence length of the wormlike chain. One can roughly say that a buckling transition appears for ratios clearly smaller than $1$. But it is crucial that one takes into account the shift of the apparent transition when a force is extracted from the onset of buckling. To illustrate the importance of thermal fluctuations we will discuss the influence they have in interpreting data from recent experiments with important biopolymers. Table \ref{tab:buckle} shows the persistence length of the $3$ polymers, ds-DNA, actin and Microtubule together with some of the typical lengths and associated transition forces. The shifted transition force is calculated from Eq. \ref{eq:transforce} with $C=1$.

The DNA tetraheda synthesized by Goodman et.al. \cite{Goodman} have sides made of double stranded DNA of a length below $10$nm. As can be read of from the table, for a lengths of $10$nm the force the structure can endure is strongly reduced by thermal fluctuations. This has to be taken into account when designing nanostructures based on DNA.

F-actin is one of the main building blocks of the cytoskeleton. It has a persistence length in the order of $9-18\mu$m \cite{Gittes1993} (the higher value is in presence of the toxin Phalloidin). actin can produce forces through polymerization. The maximum force it can produce, the stall force, was determined, by Kovar et.al. \cite{Kovar2004}, by measuring the shortest length of actin that showed buckling, when growing in between $2$ fixed points. The lengths where this was observed are given in row $2$ and $3$ of the table. The force calculation based on classical buckling considerably overestimates the force needed to buckle for the measured length since it does not take the thermal fluctuations into account.

The other important structures in the cytoskeleton are microtubules, hollow highly regular assemblies of filaments, having persistence lengths in the order of several mm's \cite{Gittes1993}, the precise value depending on several factors, like the growth speed \cite{Janson2004} and perhaps the contour length \cite{Pampaloni2006}. In buckling experiments by Janson et.al. \cite{Janson2004}, where the growth rate dependence on the applied force was studied, the lengths were such that in this case the shift by thermal fluctuations is neglegible. 
\begin{table}[t]
\begin{tabular}{l|c|c|c|c|c}
& $l_P$ &  $L$ &$h$&$\fo_c^{(0)}$ & $\fo_c$ \\
\hline
DNA &  $50$nm       &     $10$nm   &   $0.2$   &   $21$pN     &  $11$pN         \\
actin &  $9\mu$m   &     $1.2\mu$m &  $0.13$ &    $0.26$pN   & $0.16$pN      \\
actin with Phalloidin.&  $18\mu$m   &     $0.75\mu$m &  $0.04$ &    $1.3$pN   & $1.0$pN      \\
Microtubule &  $3.3$mm    &    $9.4\mu$m  & $0.028$ &   $1.5$pN     & $1.4$ pN     \\
\end{tabular} 
\caption{reduction of the force needed to buckle for some biopolymers with finite length. The reduction is calculated from Eq. (\ref{eq:transforce}) with $C=1$}\label{tab:buckle}
\end{table}

Nevertheless, the increase of thermal fluctuations when approaching buckling can also be observed in this case. These thermal fluctuations increase sharply just before buckling, followed by a strong damping of these fluctuations with increasing length (and thus increase of buckling) of the microtubule. Both these effects follow from our calculations.

The damping of the fluctuations after the onset of buckling can be inferred from the approach of the semiclassical solution towards the ``zero temperature'' classical solution. Below buckling the end point fluctuations increase from $ \vev{y^2}\cong L^2h/3 $, the classic result which follows from (\ref{eq:lowforce}), for a chain with one free end to $\vev{y^2}\approx 0.2L^2\sqrt{h}$ for an applied force corresponding to the Euler transition, as follows from (\ref{eq:forcebuck}). It should be noted that the geometry of the setup in those experiments is not immediately comparable to our calculations since the microtubule in those experiments have one end of the chain more or less hinged in a fixed position, the resulting buckling force can be up to a factor $4$ larger than in our case. Qualitatively though the results are comparable and for typical values of a persistence length of $3.3$mm and a chain length of $20\mu$m we expect the mean fluctuation of the end point to be amplified by a factor $ \approx 7$. This indeed seems to be approximately the case, although a precise analysis of their measurements is outside of the scope of this paper.

Finally, a remarkable result of our calculations is the increase of end-to-end distance by thermal fluctuations of the buckled polymers, especially in $2$ dimensions. In dense networks of actin filaments confined to the cell
cortex, the buckling is approximately $2$-dimensional. The lengthening of the buckled polymer causes then an apparent stiffening of the compressed network by the fluctuations.

\appendix
\section{elliptic functions} \label{app:elliptic}
The elliptic integrals and the Jacobi elliptic functions are functions of two variables and in the case of elliptic integrals of the third kind three. There are different equivalent choices of pairs and the choice generally depends on the situation at hand. See also \cite{Abramowitz,Gradshteyn}.

Throughout this paper we use the Jacobi form (with one exception). In that form the variables are called the \emph{argument}, $x$, and the \emph{parameter}, $m \in [0,1]$. In the literature the latter is sometimes replaced by the \emph{modulus}, $k=\sqrt{m}$. The two variables are separated by a vertical line like in $\E(x|m)$. An alternative form is the trigonometric form where the variables are the Jacobi \emph{amplitude}, $\phi=\am(x|m)$ and the \emph{modulus} $\alpha$ defined through $\sin^2(\alpha) := m$. In that case the variables are separated by a backslash. So in the notation that we use we have the elliptic integrals of the first, second and third form written as:
\begin{align}
\F(\phi|m)&=\F(\phi\backslash \alpha) & \E(x|m) &= \E(\phi \backslash \alpha) & \Pi(n;x|m) &= \Pi(n;\phi \backslash \alpha)
\end{align}
The integral of the first kind is an exception since it is in fact the inverse of the amplitude function and so $\F(x|m)$ is identical to $x$.
The complete integral of the first kind is defined as the value of $\F$ evaluated at an amplitude of $\pi/2$: $\K(m) := \F(\pi/2|m)$. The same holds for the other complete integrals, but now we can make use of the  fact that $\am^{-1}(\pi/2|m)=\K(m)$ and so:
\begin{align}
\E(m)&:= \E(\K(m)|m) & \Pi(n|m)&:=\Pi(n;\K(m)|m)
\end{align}
The double periodic Jacobian elliptic functions are defined as:
\begin{align}
\sn(x|m) &:= \sin(\am(x|m)) & \cn(x|m) &:= \cos(\am(x|m)) & \dn(x|m) := \dfrac{d}{dx}\am(x|m)
\end{align}

\section{generalized Lam\'e equation\cite{Kulic2005}} \label{app:genlam}
We are looking for a solution of the generalized Lam\'e equation:
\begin{equation}\label{eq:genLame}
\ddf{y}+p(x)y=0
\end{equation}
where $p(x) = 1+4m-\epsilon -6m\sn^2(x)$ and we are especially interested in the small $\epsilon$ limit.
The product $M(x)=y_1(x)y_2(x)$ of two solutions satisfies the third order differential equation:
\begin{equation}
\dddf{M}+4p\df{M}+2\df{p}M=0
\end{equation}
We will now construct a solution of this last equation as a series in $\sn(x)$. Write $M=\sum_{ns0} a_n\sn^n(x)$. Substitution leads to the following relation between the coefficients:
\begin{equation}
a_nm(n^3+3n^2-22n-24)+a_{n+2}(4(1+4m-\epsilon)( n+2)-(n+2)^3(1+m)) +a_{n+4}(n+4)(n+3)(n+2)=0
\end{equation}
To get a finite number of terms, the highest power has to be $4$ and we find as solution:
\begin{equation}\label{eq:functionM}
M(x)=9m^2\sn^4(x)-3m(3+\epsilon)\sn^2(x)+3\epsilon(1-m)+\epsilon^2
\end{equation}
Suppose $y_1(x)$ is one of the $2$ solutions of (\ref{eq:genLame}) that make up $M$. We can use the D'Alembert construction to get another independent solution so that $y_2$ can be written as (the Wronskian is constant):
\begin{equation}
y_2(x)=By_1(x)+Cy_1(x)\int_0^xdx'\dfrac{1}{y_1^2(x')}
\end{equation}
Using the definition of $M(x)$, and assuming $M(x)$ to be positive, we can express $y(x)$ in terms of $M(x)$ as:
\begin{equation}\label{eq:ysolution}
y=\sqrt{M(x)}\exp\left\lbrace -\int_0^x dx'\dfrac{C(m)}{2M(x')}\right\rbrace
\end{equation}
Inserting this function into the Lam\'e equation results in:
\begin{equation}
2M(x)\ddf{M(x)}-\df{M}^2(x)+C^2+4(1+4m-\epsilon-6m\sn^2(x|m))M^2(x)=0
\end{equation}
and the $C(m)$ is found by inserting the solution for $M(x)$ (\ref{eq:functionM}):
\begin{equation}
C(m)=\pm2\sqrt{\epsilon(3m(3+\epsilon)-\epsilon(1+4m-\epsilon)(3(1-m)+\epsilon))(3(1-m)+\epsilon)}\cong \pm 6\sqrt{\epsilon}\sqrt{3m(1-m)}
\end{equation}
The integrand in the exponential of (\ref{eq:ysolution}) has poles at 
$$p_{\pm} :=\dfrac{3+\epsilon\pm\sqrt{9-6\epsilon(1-2m)-3\epsilon^2}} {6m}\cong \dfrac{3+\epsilon\pm(3-\epsilon(1-2m))} {6m}$$ 
Since $\sn^2(x) \in [0,1]$, by choosing $\epsilon <0$ we can force the integrand to be regular for the parameter $m \in [0,1]$. $C(m)$ on the other hand is now imaginary , so that we have to look at linear combinations of the two solutions for a real valued solution of the homogeneous equation. Noting that $M(x)$ is now strictly negative we find as the solution with the proper boundary conditions:
\begin{equation}
y(x)= \dfrac{2\sqrt{-M(0)}}{G(m)|C(m)|}\sqrt{-M(x)}\sin\left(\dfrac{-|C(m)|}{2}\int_0^x \dfrac{dx'}{M(x')}\right)
\end{equation}
We will now evaluate the integral appearing in the sinus:
\begin{align}
\int_0^x\dfrac{dx'}{M(x')}&= \dfrac{-1}{9m^2(p_+-p_-)p_+}\int_0^x\dfrac{dx'}{1-1/p_+\sn^2(x'|m)}+\dfrac{1}{9m^2(p_+-p_-)p_-}\int_0^x\dfrac{dx'}{1-1/p_-\sn^2(x'|m)}\nonumber \\
&= \dfrac{-1}{9m^2(p_+-p_-)p_+}\Pi(1/p_+;x|m)+\dfrac{1}{9m^2(p_+-p_-)p_-}\Pi(1/p_-;x|m)
\end{align}
The elliptic integral of the third kind, $\Pi(n;x|m)$, has a behavior that depends on the value of the characteristic $n$ \cite{Abramowitz}. In our case we have characteristics $1/p_+\cong m(1-\epsilon m/3) \in (m,1)$ and $1/p_- \cong 3m/(\epsilon(1-m))<0$ both corresponding to so called circular cases.

The functional determinant is then given by:
\begin{align}\label{eq:deteps}
D_{\theta}^{\epsilon}=y(2\K(m))= \dfrac{-M(0)}{\K(m)|C(m)|} \sin\left(\dfrac{|C(m)|}{9m^2(p_+-p_-)}\left[\dfrac{1}{p_+}\Pi(1/p_+|m)-\dfrac{1}{p_-}\Pi(1/p_-|m)\right]\right)
\end{align}
where we introduced the  complete elliptic integral of the third kind, defined as usual from the incomplete one:
\begin{equation}
\Pi(n|m) := \Pi(n;\K(m) |m) = \dfrac{1}{2}\Pi(n;2\K(m)|m)
\end{equation}
The integral with the $1/p_+$ characteristic can be easily evaluated by expanding around characteristic $m$, leading to:
\begin{align}
\Pi(1/p_+|m)=\int_0^{\K(m)}\dfrac{dx}{\dn^2(x|m)}+\mathcal{O}(\epsilon)\cong \dfrac{1}{1-m}\E(m)
\end{align}
Note that we only need up to first order in $\epsilon$ and the factor multiplying the sinus in Eq. (\ref{eq:deteps}) goes as $M(0)\sim \epsilon$.

The integral with the negative characteristic can be written as an elliptic integral with positive characteristic using the following definition \cite{Abramowitz}:
\begin{align}
N_-:=\dfrac{1-p_-m}{1-p_-}\cong 1+\epsilon\dfrac{(1-m)^2}{3m}
\end{align}
resulting in:
\begin{align}
\Pi(1/p_-|m)&=\dfrac{-p_-(1-m)}{(1-p_-)(1-mp_-)}\Pi(N_-|m)-\dfrac{p_-m}{1-p_-m}\K(m) \nonumber \\
&\cong -p_-(1-m)\Pi(N_-|m)-p_-m\K(m)
\end{align}

The characteristic $N_-$ is less trivial, since we can't expand around characteristic $1$, where the elliptic integral diverges. We can express the integral in terms of yet another elliptic function, Heuman's Lambda function $\Lam_0(z|m)$, as:\cite{Abramowitz}
\begin{align}
\Pi(N_-|m)=\K(m)+\dfrac{1}{2}\pi\delta[1-\Lam_0(z|m)]
\end{align}
where 
\begin{align}
\delta&= \sqrt{\dfrac{N}{(1-N)(N-m)}}=\dfrac{1}{\sqrt{-\epsilon}}\left(\sqrt{\dfrac{3m}{(1-m)^3}}+\mathcal{O}(\epsilon)\right)\nonumber \\
z&=\sn^{-1}\left(\sqrt{\dfrac{1-N}{1-m}}\right) =\sqrt{-\epsilon}\left(\sqrt{\dfrac{(1-m)}{3m}} +\mathcal{O}(\epsilon)\right)
\end{align}
The Lambda function we can express, again following Ref. \cite{Abramowitz} in other elliptic functions as:
\begin{align}
\Lam_0(z|m)=\dfrac{2}{\pi}\left(\K(m)\E(z|1-m)-(\K(m)-\E(m))\F(\am(z)|1-m)\right)
\end{align}
Using the small argument expansion of the elliptic integrals: $\E(z|1-m)\cong  z$ we find :
\begin{align}
\Pi(N_-|m) = \K(m) - \dfrac{1}{1-m}\E(m) + \dfrac{\pi}{2(1-m)\sqrt{-\epsilon}}\sqrt{\dfrac{3m}{1-m}} +\mathcal{O}(\sqrt{-\epsilon})
\end{align}
And finally we find for the functional determinant:
\begin{align}
D_{\theta}^{\epsilon}& \cong -\dfrac{3\epsilon(1-m)}{\K(m)|C(m)|}\sin \left( \dfrac{|C(m)|}{9m(1-m)}[\K(m)(1-m)-(1-2m)\E(m)]+\pi\right) \nonumber \\
& \cong \dfrac{\epsilon}{\K(m)3m}[\K(m)(1-m)-(1-2m)\E(m)] 
\end{align}
which is Eq. (\ref{eq:thetadet})
\bibliographystyle{apsrev.bst}
\bibliography{final}

\end{document}